\documentclass[runningheads]{cl2emult}

\usepackage{makeidx}  
\usepackage{graphicx} 
\usepackage{subeqnar} 
\usepackage{multicol} 
\usepackage{cropmark} 
\usepackage{eso}      
\makeindex            

\newcommand\D{\widehat{D}}
\newcommand\order[1] { ${{\cal O}\! \left( #1 \right)}$ }
\newcommand{\eq}[1]{Eq.\ (\ref{#1})}

\newcommand{\lmax}{{\sf {L}}}

\newcommand{\etal}{{\em et al.\ }}

\begin{document}

\title*{Advanced methods for cosmic microwave background data analysis: the 
big $N^3$ and how to beat it}
\toctitle{Advanced methods for cosmic microwave background data analysis}
%
%
\titlerunning{Advanced Methods for CMB analysis}
%
\author{Benjamin D.~Wandelt}
\authorrunning{Benjamin D.~Wandelt}
%
%
\institute{Department of Physics, Princeton University, Princeton NJ 08544, USA}

\maketitle              
\begin{abstract}    In this talk I propose the first fast methods which can analyze CMB data
taking into account correlated noise, arbitrary beam shapes, non-uniform
distribution of integration time on the sky, and partial sky coverage, without
the need for approximations. These {\em ring torus methods} work by performing the
analysis in the {\em time ordered domain} (TOD) rather than on the sky map of
fluctuations. They take advantage of the simplicity of noise correlations in
the TOD as well as certain properties of the group of rotations SO(3). These
properties single out a family of scanning strategies as favorable, namely
those which scan on rings and have the geometry of an n-torus. This family
includes the strategies due to TOPHAT, MAP and Planck. I first develop the
tools to model the time ordered signal, using Fast Fourier Transform
methods for convolution of two arbitrary functions on the sphere (Wandelt and
G\'orski 2000)\cite{WandeltGorski}. Then I apply these ideas to show that in the case of a 2-torus
one can reduce the time taken for CMB power spectrum analysis from an unfeasible
order $N^3$ to order $N^2$, where $N\sim 10^5-10^8$ is the number of resolution
elements (Wandelt and Hansen, in preparation)\cite{WandeltHansen}.

\end{abstract}

\section{Introduction}

A major near-term objective in the field of Cosmology today is to
gain a detailed measurement and statistical understanding of the
anisotropies of the cosmic microwave background (CMB). While the
theory of primary CMB anisotropy is well-developed (see
\cite{review} for a review) and we are  facing a veritable flood
of data from a new generation of instruments and missions,
the single most limiting hurdle are the immense computational
challenges one has to overcome to analyze these
data\cite{borrill,Gorski,BCJK}. 

Let me outline the key problems involved. The goal is to derive two
things: an optimal sky map estimate and a set of power spectrum estimates
$\widehat{C}_l$ which, for a 
Gaussian CMB sky are a
sufficient statistic, highly informative about cosmological
parameters. In this talk I will focus on the $\widehat{C}_l$ estimation
problem, but the ideas I develop can be usefully employed for
map-making as well.

The analysis problem involves
maximizing the likelihood given the data as a function of these
$\widehat{C}_l$. 
If the observed CMB anisotropy is written as a vector $d$ the
likelihood takes the form
\begin{equation}
{\cal L}(C_l\vert d)=\frac{\exp\left[{-\frac12 {\bf d}^T ({\bf S}(C_l)+{\bf N})^{-1}{\bf d}}\right]}{\sqrt{(2\pi)^{N_{d}} \vert ({\bf
S}(C_l)+{\bf N}) \vert}},
\label{likelihood}
\end{equation}
where the signal covariance matrix $\bf{S}$ is a function of the
$C_l$, ${\bf N}$ is the noise covariance matrix and $N_d$ is the
number of entries in 
the data vector $d$. If ${\bf S+N}$ is a general matrix, evaluating
the inverse and determinant in 
this quantity  takes \order{N^3_d} operations. 

Commonly the data vector
is taken to be a set of sky map pixels, with $N_d\sim 10^5
-10^8$. Hence entries in $\bf{S}$ and ${\bf N}$ are the covariances
between  two positions on the celestial sphere. Alternatively one can
choose the spherical harmonic basis. Other bases have also been
suggested, such as  cut sky harmonics \cite{cutsky} and signal to
noise eigenmodes \cite{tegmark}, but  changing into these more general  bases
takes \order{N_d^3} operation.

Borrill (in these proceedings) presented a careful review of the 
computational tasks involved in solving this 
maximization problem in the general case, using state-of-the-art
numerical methods. The conclusion is that
in their present form these methods fail to be practical for forthcoming CMB
data sets. Given the major international effort currently under way to
observe the CMB (see, e.g.~\cite{tophat,map,planck}), finding a solution to this problem
is of paramount importance. 

The underlying reason for this failure is that the signal covariance
matrix ${\bf S}$ and noise covariance matrix 
${\bf N}$ are usually very differently structured. In  pixel space,
${\bf S}$
is  full, while ${\bf N}$  is ideally  sparse (for a well-controlled
experiment). In spherical harmonic space we are in the opposite
regime. For all-sky observations, ${\bf S}$ would be diagonal while the form
of ${\bf N}$ depends on experimental details and observational
strategy of the mission and in 
general could be any covariance matrix at all. Therefore the quantity
which enters in the likelihood 
${\bf C\equiv S+N}$ is not sparse in any easily accessible basis.

In this talk  I exhibit a solution to this
dilemma. In the
following section I will point out that modeling   
realistic observations in the time ordered domain simplifies the
problem. In particular, one can take advantage of the simple covariance
structure of the noise in the time ordered data (TOD). To be able to
formulate the likelihood problem 
in terms of the TOD one needs to be able to compute the expected signal
correlations in  the TOD. I achieve this using the 
Wandelt-G\'orski method for fast convolution on the
sphere \cite{WandeltGorski} which I briefly review in section
\ref{convolve}. In section \ref{ringtorus} I show that this implies
that the signal
correlations have a special structure for certain types of scanning strategies
namely those where the 
TOD can be thought of as being wrapped on an
n-torus. This happens to be true for the 
planned scanning strategies for TOPHAT and Planck (n=2), as well as  MAP
(n=3).

Using the results from these geometrical ideas I go on to formulate
the likelihood problem for the $C_l$ on this ring torus, with both 
${\bf S}$ and ${\bf N}$ 
sparse. In fact, they are both block-diagonal with the same pattern of
blocks; hence ${\bf C}$ is also block diagonal.

This is the first example of a fast maximum likelihood estimator which
can analyze CMB data in the presence of correlated noise. We will see
below that it can also deal with non-uniform
distribution of integration time, beam distortions,  far side lobes,
and partial sky coverage, all without needing to invoke approximations.

The exact
computational scaling of the evaluation of  ${\cal L}$ depends
somewhat on the scanning strategy used, but if the ring torus is an
2-torus it takes only \order{N^2} operations. For TOPHAT and
one of the proposed Planck scanning strategies, the ring torus is a
2-torus, for a precessing scan such as MAP's it is  a 3-torus.
For simplicity I only deal with the 2-torus
case here. Generalizations to $n\ge 3$ are trivial and they will be
discussed in \cite{WandeltHansen}.  This forthcoming paper will also
contain a presentation of the
results of applying  this method to simulated data.

\section{The Case for Analyzing CMB Data in the Time-Ordered Domain}
Why is power spectrum analysis of estimated sky maps so difficult?

\begin{description}
\item[1) Correlated noise.]
The CMB anisotropies are very small. This means that measurements
are noisy. Current detectors 
have the feature that they add correlated noise to the
signal. If we can assume that the correlated noise is  stationary and
circulant\footnote{This appears to be a good 
approximation. At least all current analyses of CMB data that contain
significant levels of 
correlated noise assume circulant or block circulant correlations!} 
then it has a very simple correlation structure in Fourier space. 
But estimating the sky from the TOD projects these simple correlations
into very complicated noise correlations between separated pixels,
which are visible as striping in the maps.

\item[2) Non-uniform distribution of integration time.] 
For most missions, practical constraints on the scanning
strategy force the available integration time to be allotted 
non-uniformly over the sky. This leads to coupling of the noise in
spherical harmonic space, in addition to coupling due to partial sky
coverage (see item 4 below). However, in the TOD,
integration time per sample is constant, by definition.

\item[3) Beam distortions and far side lobes.] 
Microwaves are macroscopic, with wavelengths of order a few $mm$ to
a few $cm$. Hence they will diffract around macroscopic objects. This
leads to distorted 
point spread functions and far side 
lobes. Simulations have to convolve these asymmetric beams with an
input sky with foreground signals varying over
many orders of 
magnitude from place to place. Analyses have to deconvolve the
observations to make accurate inferences about the underlying sky.

\item[4) Partial sky coverage.] 
The cosmological information is most visible in spherical harmonic
space. In a homogeneous and isotropic Universe, where the
perturbations are Gaussian, the power spectrum $C_l$ completely encodes
the statistical
information which is present in a map of fluctuations. Statistical
isotropy of the fluctuations around us also implies that the spherical
harmonics $Y_{lm}$ are the natural basis for representing the CMB
sky. But due to local foregrounds that need to be excised from the map or
partial sky coverage this isotropy is 
broken in actual data sets. This introduces additional signal and
noise correlations in $Y_{lm}$ space, because there is a geometrical
coupling between   $Y_{lm}$ of different $l$ and $m$.
\end{description}
There are currently no  exact and fast methods that can deal with
all of these issues.  

In order to avoid issue 1),  MAP has been designed to have very small
correlations between the noise in  separated 
pixels. A numerically exact $N^2$ maximum likelihood method
\cite{OhSpergelHinshaw} and a nearly optimal, unbiased  $N^{3/2}$
method \cite{WandeltHivonGorski} have been proposed which can deal
with issues 2) and 4) but not 3). So even without correlated noise the
results of this paper may be 
interesting for building future MAP power spectrum analysis methods,
should the issues of beam distortions and far side lobes be important. 

But both TOPHAT and Planck will produce striped
data sets beyond the capability of these or any current
algorithms. Hence, being able to deal with noise correlations is
paramount to the success of these missions. 

It is clear from 1) and 2) that the noise properties are much simpler
in the TOD than in a sky map. But what about the signal correlations?
What is the price of forfeiting the advantages of the sphere for
representing statistically isotropic signals? 
In the next section I will demonstrate that one can devise scanning
strategies which break only part of the symmetries of the sphere and
make the signal correlations in the TOD as simple as the noise properties. 

\section{Modeling the Signal in the Time-Ordered Domain Using the  Fast
Convolution Algorithm}
\label{convolve}
Recently, Wandelt and G\'orski (2000) \cite{WandeltGorski} have
introduced new methods for greatly speeding up convolutions of
arbitrary functions on the 
sphere. This reference contains a detailed description of methods which will 
enable future missions such as MAP and Planck to 
take beam imperfections into account without resorting to
approximations.  The algorithm is completely general and
can be applied to any kind of directional data, even tensor fields
\cite{CMFAWG}. I will now briefly review the ideas and quote the results.

\begin{figure}[ft]
\centering
\includegraphics[width=.8\textwidth]{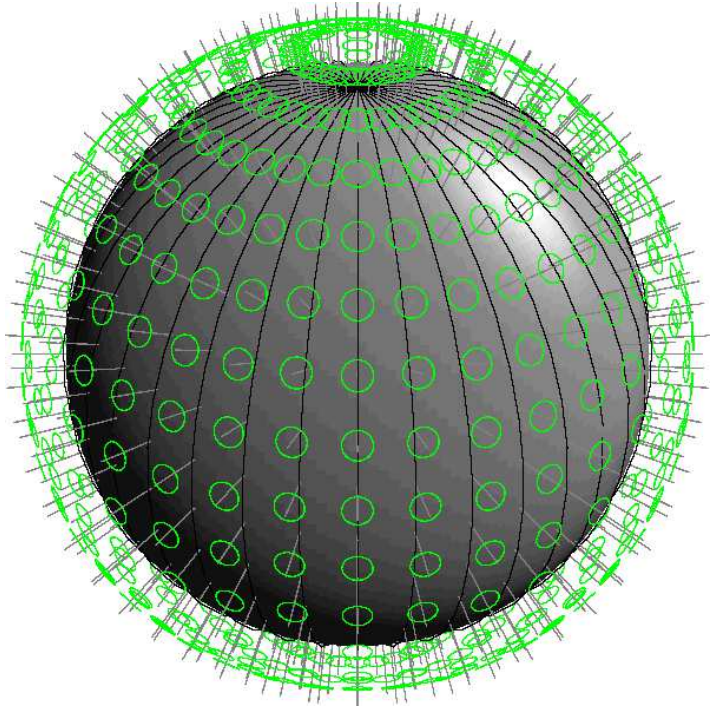}
\caption{A geometrical representation of the output of our most
general convolution method. The fact that we are dealing with
arbitrary functions means that the results are a function of the 
{\em orientation} of the beam with respect to the sky  not just the
{\em direction} it is pointing in. Therefore we need three Euler angles
to parameterize the results as shown, as rings attached to each point
on the sphere. Note that this
representation is a double cover of SO(3) because a right handed rotation
about a given direction is equivalent to a left handed rotation about
the opposite direction}
\label{fig:rotgroup}
\end{figure}

What do I mean by convolution on the sphere? Mathematically, the
convolution lives in the group of rotations SO(3), because one has to
keep track of the full relative orientation of the beam to the sky,
not just the direction it points in. To
specify the relative orientation one can 
use the Euler 
angles  $\Phi_1,\Theta$ and
$\Phi_2$.
The convolved signal for each beam orientation $(\Phi_1,\Theta,\Phi_2)$
can then be written as
\begin{equation}
T^S(\Phi_2,\Theta,\Phi_1)= \int d\Omega_{\vec{\gamma}}\,
\left\lbrack\D(\Phi_2,\Theta,\Phi_1) b\right\rbrack\!\!(\vec{\gamma})^\ast s(\vec{\gamma}).
\label{eq:start}
\end{equation}
Here the integration is over all solid angles, $\D$
is the operator of finite rotations such that $\D  b$ is the
rotated beam, and the asterisk denotes complex conjugation.
Figure
\ref{fig:rotgroup} gives a geometrical representation of the result of
such a convolution. The fact that the beam is not assumed to be azimuthally
symmetric means that at each point on the sphere, there is a ring of
different convolution results corresponding to all relative
orientations about this direction.

If $\lmax$ measures the inverse of the smallest length scale in the
smoother of the sky or beam, the numerical evaluation of  the integral in
\eq{eq:start}
takes \order{\lmax^2} operations for each tuple
$(\Phi_1,\Theta,\Phi_2)$. To allow subsequent interpolation at
arbitrary locations it is sufficient to discretize each Euler 
angle into  \order{\lmax} points and thus we have \order{\lmax^3}
combinations of them.
As a result, the total computational cost for evaluating the convolution
using \eq{eq:start} scales as \order{\lmax^{{5}}}.

\begin{figure}[ft]
\centering
\includegraphics[width=.9\textwidth]{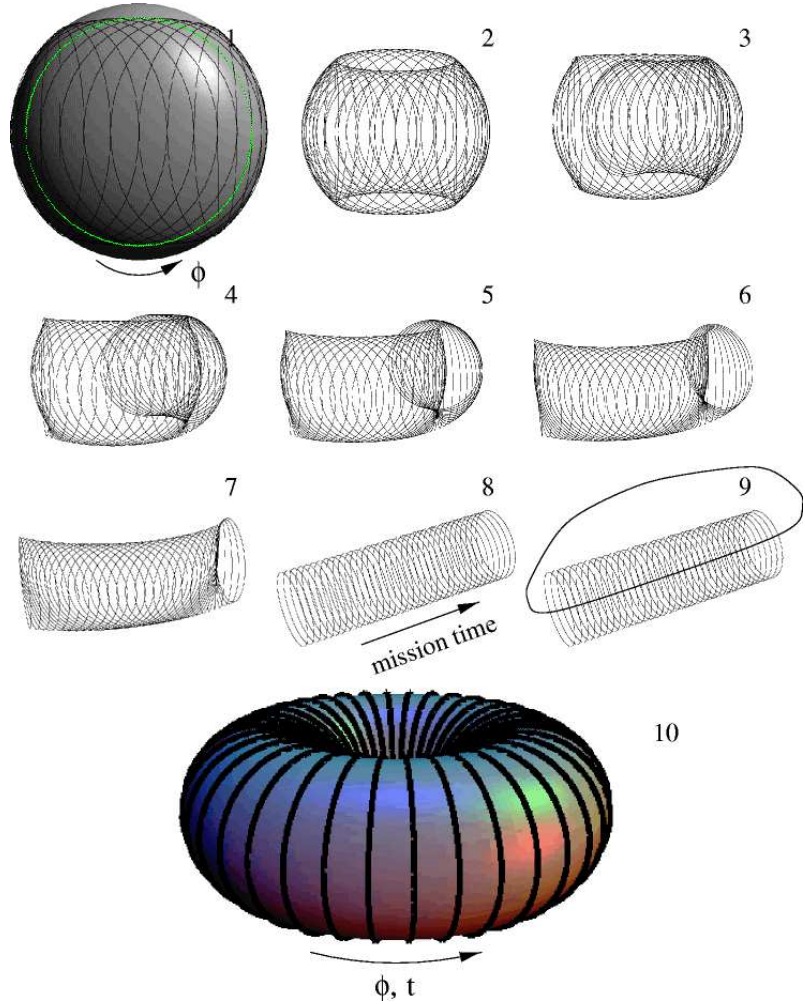}
\caption{The correspondence between a basic scan strategy (1 and 2),
the ring set (8) and a
2 dimensional ring torus (10)}
\label{fig:basictorus}
\end{figure}

It turns out that by Fourier transforming the above equation on the
Euler angles (after applying a judiciously chosen factorization) we can
reduce this scaling to \order{\lmax^{{4}}} in general. One can further
reduce the scaling to \order{\lmax^3} if one is only interested in the
convolution over a path which consists of a ring of rings at
equal latitude around the sphere. We called  such
configurations {\em basic scan paths} in reference
\cite{WandeltGorski}.
One example, covering a large part of the sphere, is shown in
Figures 2.1 and 2.2. 

Figure 2 illustrates that a basic scan path is geometrically a
2-torus and Figure 3 shows how to cut and unfold the torus for easy
visualization. 

I quote here the formula for the Fourier coefficients of the 
convolved signal on this ring torus \cite{WandeltGorski}. 
If the convolved signal on the ring torus is $T^S_{rp}$ with $r$
counting the ring number and $p$ the pixel number on the ring, then
its Fourier coefficients are
\begin{equation}
T^S_{m\,m'}(\omega=0)=\sum_{l} s_{lm} d^l_{mm'}(\theta_E)  X_{lm'}.
\label{eq:specialresult}
\end{equation}
Here the $s_{lm}$ are the spherical harmonic multipoles for the CMB
sky and $\theta_E$ fixes the latitude of the spin axis on the
sky. The quantity  
\begin{equation}
X_{lm}\equiv \sum_{M} d^l_{mM}(\theta)b_{lM}^\ast
\label{eq:precompute}
\end{equation}
is just the rotation  of the beam multipoles $b_{lm}$ for an
arbitrary beam by $\theta$, the opening angle of the scan
circle. $X_{lm}$ can be precomputed. Counting indices in 
\eq{eq:specialresult} confirms
that computing the convolution along a basic scan path takes only
$\sim\lmax^3$ operations. 

\begin{figure}[ft]
\centering
\includegraphics[width=\textwidth]{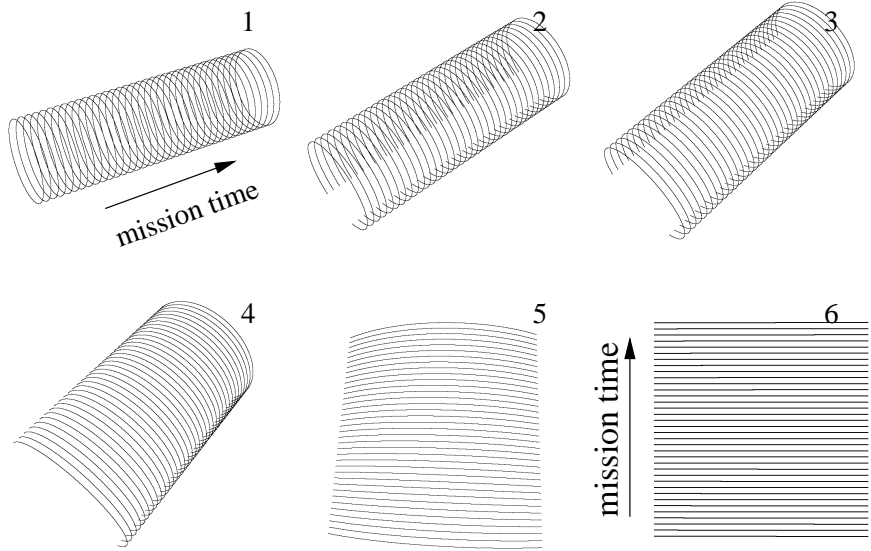}
\caption{Here I show continuing from Figure 2.8 how to present the
convolved signal in 
the ring set (1) as a two dimensional density plot where the rings are
cut, extended into lines, and stacked vertically (6)}
\label{fig:sphere}
\end{figure}

\begin{figure}[ft]
\centering
\includegraphics[width=\textwidth]{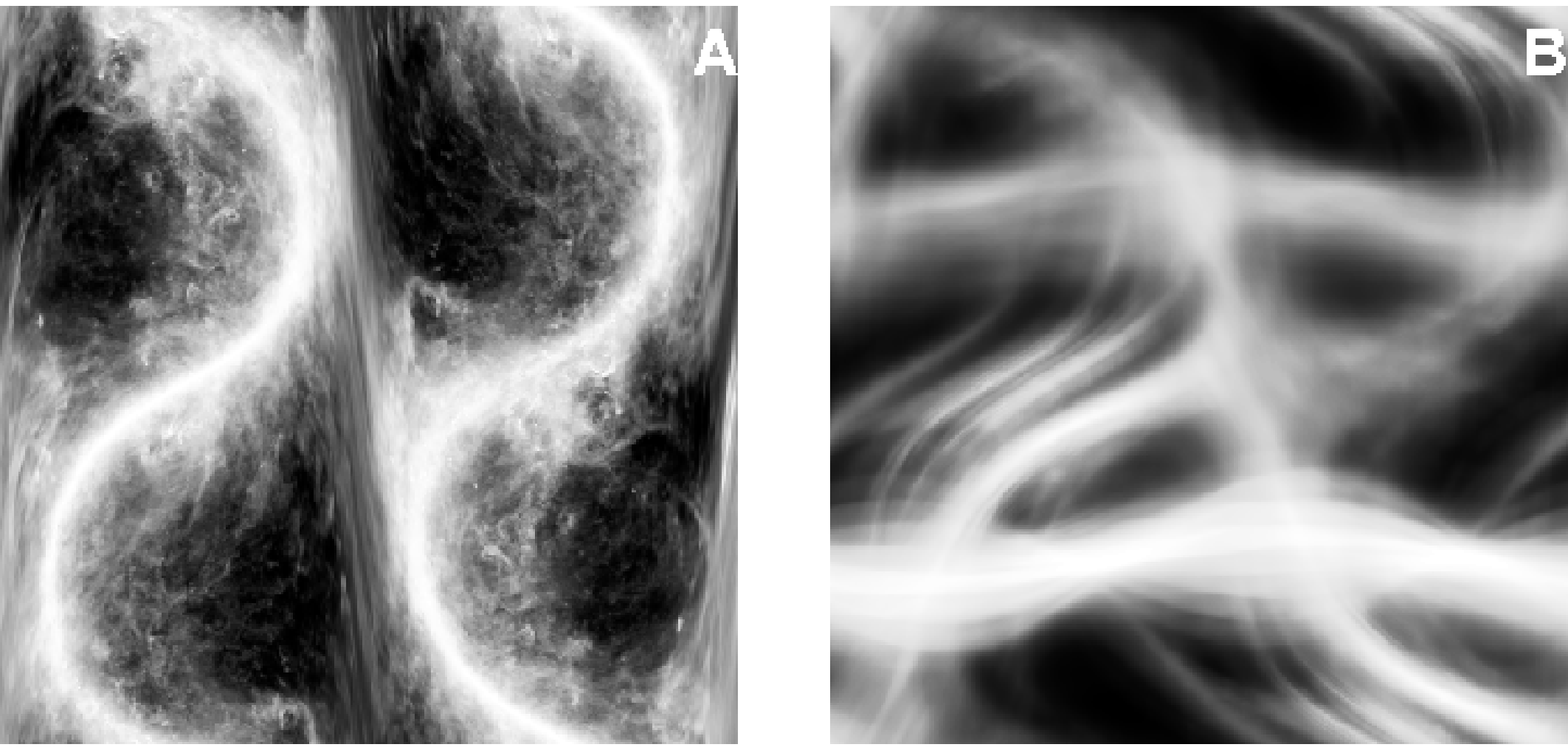}
\caption{Here I show the ring set that  results from the fast
convolution algorithm along a basic scan  path in the representation
shown in Figure 3.6. In 
both (A) and (B) I used the Schlegel Finkbeiner Davis  model of dust
emission \protect\cite{SFD98} at 100GHz, sampled on $\sim 10^7$ pixels and the worst Plank
LFI horn  with a main beam size of about
30 arcminutes FWHM. Part (A) shows the convolution using only the 5
degrees of the  
beam around the main beam and part (B) shows the
convolution with the full far side lobe pattern, excluding the inner 5
degrees. The beam and sky vary over many orders of magnitude. In (B)
even discreteness effects of the ECP 
pixels which represented the beam side lobes can be seen}
\end{figure}

Figure 4 shows
the results of such a convolution for the main beam 
(inner 5 degrees) and
the far side lobes (everything except the inner 5 degrees) of the
physical model of one Planck LFI 30GHz 
horn. The nominal resolution of this horn is 30 arcminutes FWHM. This
computation of $4\times 10^6$ convolved samples from the Schlegel 
Finkbeiner Davis  dust emission model \protect\cite{SFD98} at 100 GHz ($12\times 10^6$
pixels) took $\sim 15$ minutes. That compares with a projected time of
several months if the integrals had been done in pixel space, even
when using efficient grids of varying resolution.

\section{Ring Torus Analysis: Beating the Big $N^3$}
\label{ringtorus}
Using the results from the previous section we can now write down an
statistical model of the 
time-ordered data as a function of the $C_l$ and experimental
parameters. 

I will first derive the signal-correlation matrix  as a function of
the $C_l$ and then
go on to derive the noise correlation as a function of instrumental
parameters (like the shape of the noise power spectrum in the TOD) and
the scanning strategy (in this case the dimensions and position of the
basic scan path).
Detailed derivations  can be found in \cite{WandeltHansen}. 

\subsection{Signal Correlations}

What we are interested in are the correlation properties of the signal
on the ring torus, \eq{eq:specialresult}. We can now easily compute
the correlation matrix  
\begin{eqnarray}
{\cal T}^S_{m\,m'\;M\,M'}&\equiv& \left\langle T^S_{m\,m'}
T^S_{M\,M'}\right\rangle\\
&=& \fbox{$\delta_{mM}$}\sum_{l} C_{l} d^l_{mm'}(\theta_E)  X_{lm'}d^l_{MM'}(\theta_E)
X_{lM'}.
\end{eqnarray}
The boxed Kronecker delta shows that the  correlation matrix in
Fourier space on the ring torus is {\em block diagonal}. 

\subsection{Noise Correlations}

I start by assuming that we can model the noise in the time ordered
domain as a stationary
Gaussian process whose Fourier components $\widetilde{T}^N_k$ have the
property
\begin{equation}
\left\langle\widetilde{T}^N_k\widetilde{T}^{N\ast}_{k'}\right\rangle =
P(k)\delta_{kk'}.
\end{equation}
The noise power spectrum in the time-ordered domain is well described
for practical purposes as $P(k)=\sigma^2[1+(k_{knee}/k)^\alpha]$, with $\alpha\in]0,2]$.

In the following I will allow for the complication that  the scanning
strategy involves integrating repeatedly on 
a fixed ring then the signal remains the same for each scan of
that ring. One can just co-add all scans over the
same ring without losing information. 

To be specific, if we define $N_R$ to be the number of pixels
per ring,  $N_s$  the
number of times the satellite spins while observing a fixed ring, and $N_r$
to be the number of rings for a one-year mission, we
can write $N_{TOD}= N_R N_s N_r$ for the number of samples in the
TOD.   In the following I will choose $N_r=N_R$ 
for simplicity, though this is not required by the formalism and can
be generalized trivially. The ring torus will then have $N_r^2$ elements.

These properties of the noise
covariance in the time-ordered data allow writing the noise
covariance on the ring set as \cite{WandeltHansen}
\begin{equation}
\left\langle{T}^N_{rp}{T}^{N\ast}_{r'p'}\right\rangle\equiv C(r-r',p-p'),
\label{first}
\end{equation}
where $C$ is periodic with period $N_r$ in its first entry;
letting $\Delta\equiv r-r'$ we have
\begin{equation}
C(\Delta ,p-p')=C(\Delta +j N_r,p-p'), \quad j\in {\cal Z}.
\label{circulant}
\end{equation}

This implies for the covariance of the
Fourier components of the 
the co-added noise on the rings
\begin{eqnarray}
{\cal T}^N_{m'mM'M}&\equiv&
\left\langle\widetilde{T}^N_{m'm}\widetilde{T}^{N\ast}_{M'M} 
\right\rangle\label{noisecovariance}\\
&=&\frac{1}{(N_r)^3}\fbox{$\delta_{mM}$}\sum_{p,p'=0}^{N_r-1}e^{-\frac{2\pi
i}{N_r}  (m'p-M'p')} \sum_{\Delta =0}^{N_r-1}e^{-\frac{2\pi i}{N_r}  m
\Delta }C(\Delta ,p-p')  \nonumber
\end{eqnarray}
where I used (\ref{circulant}) to
obtain the boxed Kronecker delta which flags that the noise
covariance is {\em block diagonal} on the ring torus in exactly the
same way as the signal.

\subsection{From $N^3$ to $N^2$: Why it Works and Other Advantages}

I just showed that by choosing the scanning strategy such
that the TOD can be wrapped (co-added) onto a  2-torus one can find a
basis  in which both noise and signal are block diagonal and hence
sparse. What is more, the data can be easily transformed into this basis
just by computing the Fast Fourier Transform of the co-added TOD.

Why does it work? The ring
torus exhibits a spatio-temporal symmetry: time proceeds linearly
around the torus, (``$t$'' in  Figure 2.10), in the same direction as the
azimuthal angle (``$\phi$'' in Figure 2.1 and 2.10). Stationarity of the
noise in the TOD and isotropy of the signal on the sphere are of
course broken on the torus, but both show up as a ``partial isotropy''
(only in the azimuthal direction), which expresses itself as a
simplified correlation structure in Fourier space.

One can therefore write down the likelihood \eq{likelihood} on the ring torus
by simply substituting
\begin{equation}
d_{\rm map}\rightarrow d_{\rm ring~torus},\quad {\bf
S}\rightarrow{\cal T}^S,\quad
{\bf N}\rightarrow{\cal T}^N 
\end{equation}

Due to the block diagonality, the inverse and determinant in the
\eq{likelihood} 
only take \order{N^2} operations to 
evaluate. The gradient is similarly easy to evaluate
\begin{equation}
-2\frac{\partial\ln {\cal L}}{ \partial C_l}=
Tr({\bf C}^{-1} {\bf C,}_{C_l})- d^T{\bf C}^{-1} {\bf C,}_{C_l}{\bf C}^{-1}d
\end{equation}
which only takes \order{\lmax^3\sim N^{3/2}} operations to compute for
each component.

Other than speed there are other advantages of ring torus methods.
First, this approach bypasses map-making on the sphere for the purposes
of power spectrum estimation.  Not having to make maps first, helps avoiding
questions like ``What does my favorite de-striping algorithm do to the
signal?''. On the ring torus, one fully models the striping due to 
correlated noise and hence does not rely on  such {\em ad hoc} methods. 

While the sky map is not necessarily the best
starting point for the purposes of $C_l$ estimation, one should
of course make the best map one possibly can in order to identify
systematic errors {\em  etc.}. Reference \cite{WandeltGorski} shows
how the ring torus speeds up iterative methods for deconvolving with
asymmetric beams. The block diagonal noise correlations on the torus
mean that one can use the same geometrical ideas for solving the map
making equations in the presence of noise.

In fact both signal and noise
covariances are sparse in this basis, hence
this method can be used to quickly Wiener filter the data on the ring torus
once the $C_l$ are known to optimally estimate the map of CMB
fluctuations. 

\section{Possible Extensions and Conclusions}

In this talk I constructed  for the first time an exact formulation of
the maximum likelihood problem of power spectrum analysis on the
sphere which can deal with correlated noise, realistic beams, partial
sky coverage, and non-uniform noise with a computational scaling of
\order{N^2}$\!\!$. This is a significant advance over other available exact
methods with the same scaling, all of which assume white noise and
azimuthally symmetric beams amongst other assumptions.

But however more realistic this approach is, real observations will
still be more complicated. In the case of nearly all sky observations, I
have not showed how to deal with small areas where foregrounds
dominate such as dust emission from the galactic plane,  or bright
point sources. Also, irregularities
in the scanning strategy will somewhat spoil the symmetry of spatial
correlations in the signal and temporal correlations in the noise
which the ring torus exploits.

While it is hopeless to attempt to address these further complications
using current methods, I hope that ring torus methods provide a new
avenue of investigation. The idea is that perturbation theory is
powerful but only works when exact solutions are known. So I propose
to use exact inverses, computed using the methods I presented in this 
talk as preconditioners to speed up conjugate
gradient methods for computing $C^{-1}d$ for more general cases, e.g.
galactic cuts, point sources and imperfect scanning. This is 
analogous to the approach of the authors of \cite{OhSpergelHinshaw}
who built numerical tools for the MAP analysis by perturbing around an exact
solution for azimuthal sky cuts.  

Under the stated assumptions I have shown that the power spectrum
analysis problem simplifies 
greatly when it is considered in the domain of co-added time-ordered
data which has the geometry of a torus. Extensions of the formulae
I presented here to more 
general scanning strategies (e.g. a precessing scan, such as MAP's or
another proposal for Planck) are easy to derive  using the more general 
results of \cite{WandeltGorski}. In this case the efficiency depends on the
size of the extra dimensions which the torus acquires. 

If beam distortions and far side
lobes are an issue, but noise correlations are not, then
it may be more efficient to perform the analysis on a 3 torus
which does not follow the scanning strategy, but parameterizes the
full group of rotations SO(3) of the beam with respect to the sky and uses
the {\em total convolution} formula mentioned in 
\cite{WandeltGorski}. This still allows using FFTs, but takes
advantage of the fact that azimuthal beam asymmetries are usually
mild. This returns an \order{N^2} method  with a prefactor depending
on the number of harmonics needed to model the azimuthal beam
asymmetry. Note that in this case the scanning strategy is not
constrained  --- but it should preferably visit every point on the sky
and cover it with many different beam orientations so the group
manifold of SO(3) is covered densely.

Once it is possible to deal with correlated noise, the fear of
striping that has haunted the field is mitigated. Imagine that 
given a fixed rms noise amplitude we could choose how much
of it we would like to be white and how much we would like to be
correlated. What should we wish for? Actually, we should wish for maximally
correlated noise! Correlated noise has regularity in it,
which can be modeled, and hence removed as long as it does not mimic
the signal we want to measure. White noise can never be
removed, only at the expense of smoothing out structure in the signal.

I am grateful to the conference organizers for an excellent meeting,
A.~J.~Banday, K.~G\'orski, F.~Hansen  for valuable discussions and
enjoyable collaborations and to the NASA MAP/MIDEX program for support.

\end{document}